# An Analytical Approach for Project Managers in Effective Defect Management in Software Process

T.R. Gopalakrishnan Nair
Advanced Software Engineering Research Group
Research and Industry Incubation Centre
Dayananda Sagar Institutions
Bangalore, India
trgnair@ieee.org
trgnair@gmail.com

V. Suma
Advanced Software Engineering Research Group
Research and Industry Incubation Centre
Dayananda Sagar Institutions
Bangalore, India
sumavdsce@gmail.com

N. R. Shashi Kumar
Advanced Software Engineering Research Group,
Research and Industry Incubation Centre
Dayananda Sagar Institutions
Bangalore, India
nrshash@gmail.com
chandravalli@yahoo.com

*Abstract:* Defect estimation and prediction are some of the main modulating factors for the success of software projects in any software industry. Maturity and competency of a project manager in efficient prediction and estimation of resource capabilities are one of the strategic driving forces towards the generation of high quality software. Currently, there are no estimation techniques developed through empirical analysis to evaluate the decision capability of a project manager towards resource allocation for effective defect management. This paper brings out an empirical study carried out in a product based software organization. Our deep investigation on several projects throws light on the impact of decision capability of project manager towards accomplishment of an aforementioned objective. The paper enables project managers to gain further awareness towards the significance of predictive positioning in resource allocation in order to develop high quality defect-free software products. It also enhances the maturity level of the company and its persistence in the competitive atmosphere.

*Keywords: Defect Management, Project Manager, Inspection, Testing, Software Engineering.*

## I. INTRODUCTION

The development and use of software continue to be one the key technologies for the success in any business domain. Way back half a century, no one could have predicted that the software would become an indispensable technology for business, science and engineering. Advancement in software has enabled the creation of new technologies like genetic engineering, nano technology, telecommunication etc. Software is also one of the most vital components in embedded technologies such as medical, automobiles, military, industrial, entertainment etc. Due to the varied significance of software, it is important for the software community to develop high quality software products with optimized time, cost and resource in order to attain complete customer satisfaction [1].

Additionally, high quality software is reliable, dependable and is defect-free. A defect in an application can lead to a harmful situation at all the phases of software development. Anything related to defect is not a state and is a continual process. Owing to the magnification and propagation nature of software defects, effective defect management deems to be an elementary and significant activity during software development [2] [3].

Since, a project manager has an overall responsibility for the successful initiation, planning, design, execution, monitoring, controlling and closure of a project or a part of a project, he has a considerable role in achieving high quality software. Thus, the success or failure of the project depends upon the competency of a project manager in realizing right estimation, prediction and evaluation of resources required for the development of the project [4].

This paper therefore presents an empirical investigation on several projects from a product based software industry to signify the role of project manager in achieving effective defect management. Section II of the paper briefs about the background work for this investigation. Section III presents the empirical analysis of several projects through a case study. Observations and their inferences drawn from this deep investigation are presented in Section IV. Similarly, Section V provides the summary of this paper.

## II RELATED WORK

Success of any project depends on several factors such as scheduled time, cost, number of developers in the team, experience level of the team members, domain knowledge, choice of technology and implementation of standards etc. Since, a project manager is responsible for resource planning of projects, accurate estimation and prediction of resources contributes for the success of the project. Therefore, efficiency level of project managers influences the success level of a project. Authors in [5] express that the personality of a project manager acts as a good predictor for the success of projects. They recommend the big five personality traits namely Extraversion, Openness, Conscientiousness, Neuroticism, Agreeableness as modulating factors for the success of the project. They further emphasize on the lack of research in the area of role of project manager towards the success of the project. Our investigation focuses on this arena. However, this research indicates a vital need for analytical reasoning from project managers towards effective resource allocation for defect management in order to realize successful software projects [5].

Authors in [6] feel that the knowledge of project manager plays a vital role in the success or failure of projects. They further express that an experienced Project Manager with his skill set comprising of integration, scope, time, cost, quality, human resource, communication management, risks and procurement management influences the success of the project. Our investigation therefore provides an empirical study of several projects to bring in awareness of the project manager's role in effective defect management which is one of the influencing factors for the success of a project [6].

Authors in [7] state that software quality also depends upon the team satisfaction. They state that product quality was low with moderate team satisfaction and that with an increased team satisfaction, the quality of the product further accelerates [7]. Further, authors in [8] suggest that there is also a role of culture on the evolution of group potency among project team members. They state that project team potency is influenced by project team culture and that the project success and project member satisfaction is influenced by project team potency [8]. Further, defect management is one of the core demands for the success of the project and it is a fact that team performance influences the effective defect management [9] [10]. Since, project manager is responsible for planning and scheduling of team, our deep investigation on several projects revealed the role of project managers to achieve effective defect management through right selection of team and its skills.

Author in [11] states that certain critical factors such as project management process, project communication, project participators, collaboration and information sharing mechanisms etc. influences the success of the project. They further emphasize on the necessity of project managers to enhance their capabilities in terms of project management skills, depth of management knowledge, ability to solve problems in practice and to enhance the team ability to deal with the changes in order to achieve better project management [11].

Nevertheless, effective defect management is one of the critical factors for the success of the project [12]. Our investigation through a case study on several projects reveals the significance of project manager.

III   CASE STUDY

This paper presents a case study of a product based software industry. The company functions on business intelligence (BI). Its expertise is based on years of implementation of complex business intelligence solutions across diverse technology. The company is partnered with the world's leading data warehousing technology providers.

This work includes study of several projects developed in the company. The scope of investigation includes projects developed from the year 2001 onwards and up to 2011. In order to resolve the varied complexities of production, this study considers two categories of projects namely medium category of projects and large category of projects. Medium projects require less than 5000 hours of software development time. Large projects consider more than 5,000 hours of software development time. These projects were developed on Oracle database and used Java based tools in Linux Operating system environment. The data related to this investigation is collected from the Document Management Repository of the company. The work focuses on the three phases of software development namely requirement analysis phase, design phase and implementation phase. This paper presents a selection of 6 projects which are sampled from both categories of projects.

Table 1 shows the sampled data at requirements analysis phase. Table 2 and Table 3 illustrate the sampled data at the design and implementation phase of the company under study. They provide the information about the log of defects and defect capturing status of the company in terms of defect management strategies followed in the company.

The tables in turn reflect the project manager's role in planning and scheduling of resources which also includes time, number of resource personnel and their experience level.

Fig. 1 and Fig. 2 show the number of defects detected by inspection team and test team at various phases of software development respectively.

Table 4 shows the results as obtained for different projects. In this table, we have computed X, the total number of defects estimated, Y, the total number of defects detected, T, the total number of defects actually present in all the phases for 6 sampled projects, Z, the total number of defects un-captured in all the phases of the sampled 6 projects. From the aforementioned computations, it is now possible to evaluate Project Manager Efficiency Level (Ep) and Prediction Efficiency PE ($A_i$) of a project manager in terms of defect estimation which is illustrated through the sampled 6 projects.

## IV OBSERVATIONS

From the empirical data, it is observed that with increase in performance of team at inspection process, the test effort decreases [2]. This investigation focuses on the significance of role of project manager in effective defect management in order to achieve quality and success of the project.

Table-1: The sampled data at requirement phase

| PROJECTS | P1 | P2 | P3 | P4 | P5 | P6 |
|---|---|---|---|---|---|---|
| Total project time in hours | 2400 | 2880 | 4320 | 4800 | 11520 | 11520 |
| Requirements analysis time | 400 | 80 | 320 | 240 | 560 | 960 |
| Inspection time scheduled | 16 | 32 | 32 | 32 | 64 | 80 |
| Number of inspectors involved | 2 | 2 | 2 | 2 | 2 | 2 |
| Experience level of inspectors (years) | 8 | 8 | 8 | 8 | 10 | 10 |
| Defect count estimation | 16 | 18 | 32 | 26 | 40 | 48 |
| Number of defects detected | 14 | 12 | 30 | 23 | 36 | 32 |
| Defects actually captured | 14 | 10 | 30 | 24 | 35 | 35 |
| Number of defects not captured | 0 | 2 | 0 | 2 | 5 | 13 |
| Defects due to bad fixes | 4 | 4 | 4 | 6 | 11 | 7 |
| Testing time scheduled | 28 | 33 | 50 | 55 | 132 | 132 |
| Number of testers | 1 | 1 | 1 | 1 | 2 | 2 |
| Experience level of testers (years) | 5 | 5 | 5 | 5 | 5 | 8 |
| Defect count estimation | 12 | 6 | 15 | 26 | 26 | 48 |
| Number of defects detected | 12 | 6 | 15 | 23 | 24 | 32 |
| Defects actually captured | 12 | 3 | 15 | 24 | 28 | 35 |
| Number of defects not captured | 0 | 0 | 0 | 0 | 2 | 13 |
| Number of defects due to bad fixes | 2 | 4 | 2 | 6 | 3 | 7 |

Table-2: The sampled data at design phase

| PROJECTS | P1 | P2 | P3 | P4 | P5 | P6 |
|---|---|---|---|---|---|---|
| Total project time in hours | 2400 | 2880 | 4320 | 4800 | 11520 | 11520 |
| Inspection time scheduled | 480 | 576 | 864 | 960 | 2304 | 2304 |
| Number of inspectors involved | 1 | 1 | 1 | 1 | 2 | 2 |
| Experience level of inspectors (years) | 5 | 5 | 5 | 5 | 5 | 8 |
| Defect count estimation | 22 | 28 | 28 | 22 | 56 | 32 |
| Number of defects detected | 20 | 27 | 26 | 20 | 36 | 35 |
| Defects actually captured | 24 | 32 | 26 | 18 | 35 | 33 |
| Number of defects not captured | 0 | 0 | 0 | 2 | 1 | 2 |
| Defects due to bad fixes | 4 | 4 | 2 | 6 | 11 | 7 |
| Testing time scheduled | 120 | 144 | 216 | 240 | 576 | 576 |
| Number of testers | 1 | 1 | 1 | 1 | 2 | 2 |
| Experience level of testers (years) | 8 | 8 | 8 | 8 | 8 | 10 |
| Defect count estimation | 14 | 12 | 12 | 28 | 24 | 32 |
| Number of defects detected | 12 | 12 | 10 | 20 | 23 | 35 |
| Defects actually captured | 12 | 16 | 8 | 26 | 22 | 32 |
| Number of defects not captured | 0 | 2 | 2 | 2 | 1 | 3 |
| Number of defects due to bad fixes | 4 | 4 | 4 | 3 | 3 | 7 |

Table-3: The sampled data at implementation phase

| PROJECTS | P1 | P2 | P3 | P4 | P5 | P6 |
|---|---|---|---|---|---|---|
| Total project time in hours | 2400 | 2880 | 4320 | 4800 | 11520 | 11520 |
| Inspection time scheduled | 840 | 1008 | 1512 | 1680 | 4032 | 4032 |
| Number of inspectors involved | 5 | 5 | 5 | 5 | 5 | 8 |
| Experience level of inspectors (years) | 3 | 3 | 3 | 3 | 3 | 3 |
| Defect count estimation | 24 | 34 | 26 | 32 | 42 | 43 |
| Number of defects detected | 22 | 32 | 24 | 30 | 40 | 42 |
| Defects actually captured | 22 | 36 | 24 | 28 | 38 | 40 |
| Number of defects not captured | 0 | 0 | 0 | 2 | 4 | 2 |
| Defects due to bad fixes | 4 | 4 | 2 | 4 | 11 | 4 |
| Testing time scheduled | 288 | 346 | 518 | 576 | 1382 | 1382 |
| Number of testers | 2 | 2 | 2 | 2 | 2 | 2 |
| Experience level of testers (years) | 4 | 4 | 4 | 4 | 4 | 8 |
| Defect count estimation | 16 | 12 | 12 | 12 | 8 | 46 |
| Number of defects detected | 14 | 10 | 10 | 10 | 8 | 44 |
| Defects actually captured | 12 | 10 | 8 | 10 | 8 | 42 |
| Number of defects not captured | 2 | 0 | 2 | 0 | 2 | 2 |
| Number of defects due to bad fixes | 4 | 4 | 4 | 2 | 3 | 5 |

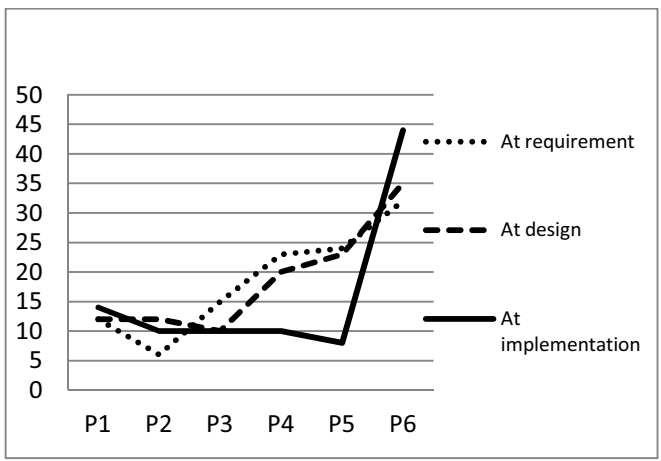

Figure 1: No. of defects detected by inspection team at three phases

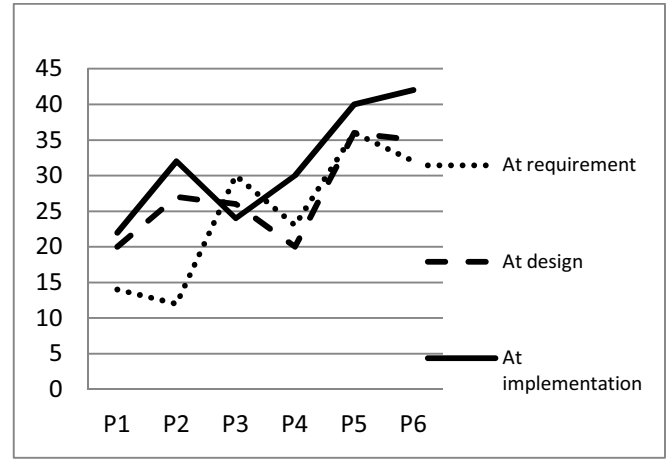

Figure2: No. of defects detected by testing team at three phases

Table-4: The results as obtained for different projects

| Projects | P1 | P2 | P3 | P4 | P5 | P6 |
|---|---|---|---|---|---|---|
| Total defects estimation (X) | 104 | 110 | 125 | 146 | 196 | 249 |
| Sum of defects detected(Y) | 96 | 107 | 111 | 130 | 166 | 217 |
| Total defects present (T) | 116 | 123 | 133 | 153 | 209 | 257 |
| Total defects not captured ( Z) | 20 | 16 | 22 | 23 | 43 | 40 |
| Project Manager Efficiency Level (Ep) | 92.31 | 97.27 | 88.80 | 89.04 | 84.69 | 87.15 |
| Prediction Efficiency, PE(Ai) = X/T | 89.66 | 89.43 | 93.98 | 95.42 | 93.78 | 96.89 |
| Average Prediction Efficiency (Avg.PE ($A_i$)) | 93.19 | | | | | |

The tables 1 through 3 indicate the defect capturing ability of the inspection and test team as assigned by the project manager. From the emperical investigations, following inferences can be drawn. Let X be the total number of defects estimated at all the phases. Let Y be the total number of defects detected by both inspection and test team at all the phases of software development. Let T be the total number of defects present which includes detected defects and defects introduced due to badfixes during the development process. The aforementioned analysis is depicted below as equations numbered (1) through (7). The variable n represents phases which is equal to 3 namely requirements phase, design phase and implementation phase in this investigation.

$$X = \sum_{i=1}^{n}(No. of\ Defects\ Estimated\ in\ all\ the\ phases) \quad (1)$$

Fig. 3 shows X, the total number of defects estimated according to the above specified equation.

$$Y = \sum_{i=1}^{n}[Defects\ detected\ by\ the\ Inspection\ Team] + \sum_{i=1}^{n}[Defects\ detected\ by\ the\ Test\ team] \quad (2)$$

Fig. 4 indicates Y, the total number of defects detected.

$$T = \sum_{i=1}^{n}[No. of\ Defects\ detected + No. of\ Defects\ introduced\ due\ to\ bad\ fixes] \quad (3)$$

Fig. 5 shows T, the total number of defects totally present at all the phases of the sampled projects

$$Z = [T-Y] \quad (4)$$

Z, the total number of defects not captured by both teams is shown in the Fig. 6

Effectiveness of the team performance depends upon time scheduled, level of experience and skill of the team members[13][14]. Therefore, it is imperative for the success of the project to have proper planning of resources. Since, project manager is responsible for planning of resources, the success of the project and the success of the company depends upon project manager also. Further, effective defect managmenet is one of the crietia for the success of the project, it is now inevitable for the company to evlauate the efficieny level of project manager in terms of effective defect managment very specifically. Let efficiency level of project manager in terms of defect management be Ep. Let N be the total number of projects which are planned, scheduled, controlled and monitored by the project manager. The sampled projects considered in this paper areoperated by the same project manager.

$$Ep = \sum_{i=1}^{N}((X-Y)/X) * 100 \quad (5)$$

Fig. 7 shows Ep, the Project Manager Efficiency Level for the 6 sampledprojects.

Success of the project depends upon several factors. The factors deal with the role of project manager in effective planning, scheduling, monitoring, controlling and closure of the project. Therefore, project manager's efficiency level in effective defect management is one of the modulating factors for the success of the project. Let A, be the success level of the project and $a_0$ to $a_n$ represent several influencing factors of the project manager such as personality, project management skills, depth of management knowledge, ability to solve problems in practice etc.

$$A = f(a_0+a_1+a_2+\ldots \ldots +a_n) \quad (6)$$

Where a=Ep in different domains

However, the success level of the company depends upon project success and the project success further depends upon several factors such as effective defect management, development of high quality product within scheduled time, cost, number of developers in the team, experience level of the team members, domain knowledge, choice of technology and implementation of standards etc. along with the efficiency level of project manager. Let $C_s$ be the success level of company which is proportional to the success level of project

$$Cs = K\ A \quad (7)$$

where K is a success co-efficient which is not presented in this paper.

The Predicting efficiency PE ($A_i$) of a project manager is further plotted in the Fig. 8.

Thus, from the above approaches, it is apparent that awareness of project manager's efficiency level is one of the basic requirements for the success of the project. It is evident that, any flaw in that can lead to process deficiency and subsequently it can lead to loss in quality and business. Currently, the role and responsibilities of a project manager are estimated mostly on intuitive analysis. This paper throws light upon the need for a development of analytical model for the project manager's efficiency in order to enhance their evaluation and prediction capabilities which affect the project's success in terms of quality and cost. This study is an attempt to bring a special awareness on the significance of a project manager in effective defect management which is one of the primary controlling factors to achieve high quality software in order to sustain the competitive industrial environment. Even though other capabilities of project manager like skills for follow-up, tracking, teamwork etc. is appreciable, one thing needs to be emphasized here is that the low efficiency in prediction of defect occurrence can affect the results creatable through the skills cited above.

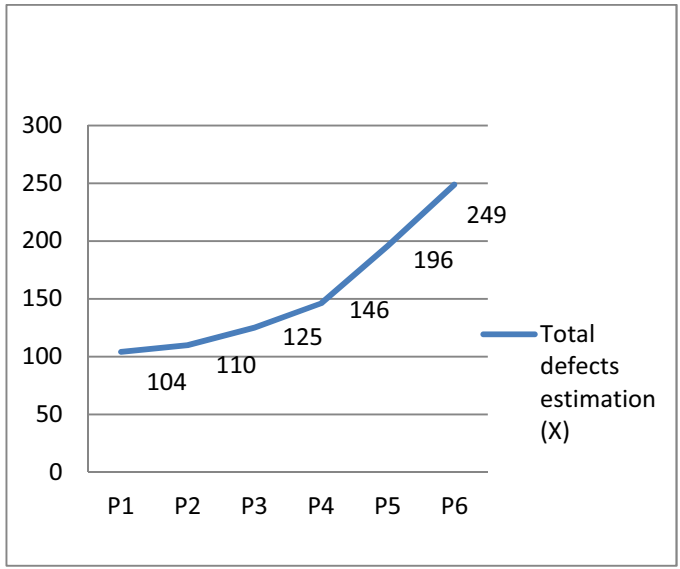

Figure 3. Total number of defects estimated (X)

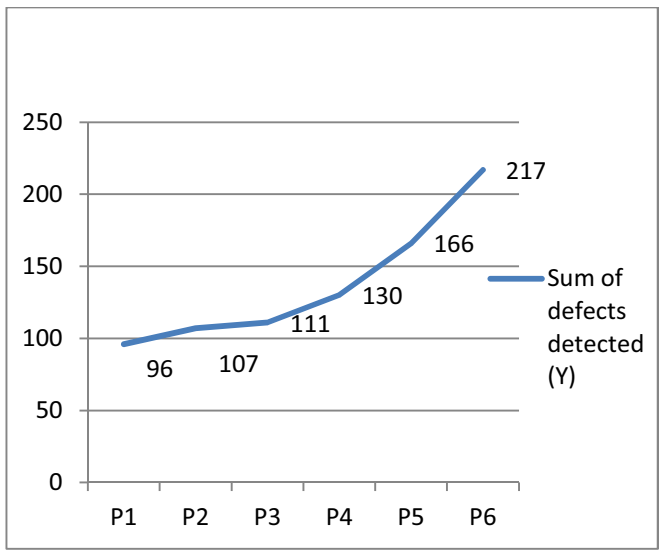

Figure 4. Total number of defects detected (Y)

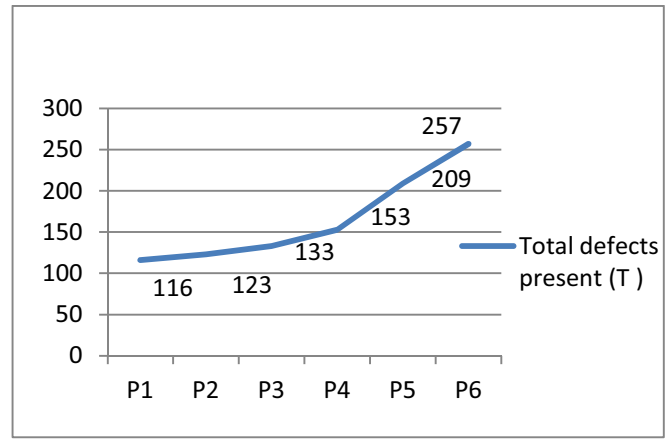

Figure 5. Total number of defects present (T)

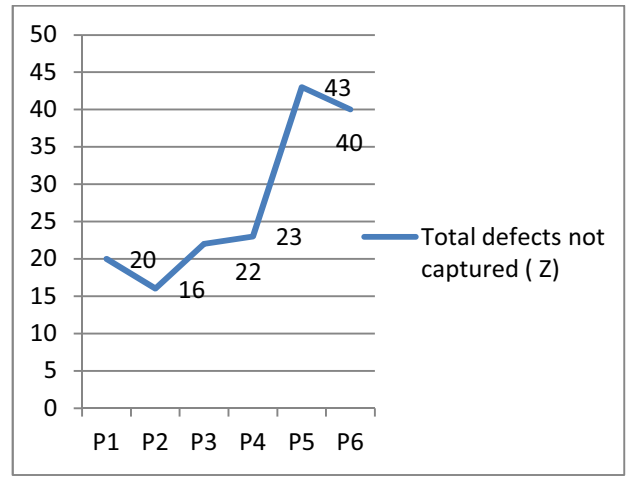

Figure 6. Total number of defects not captured (Z) at the three phases of software development

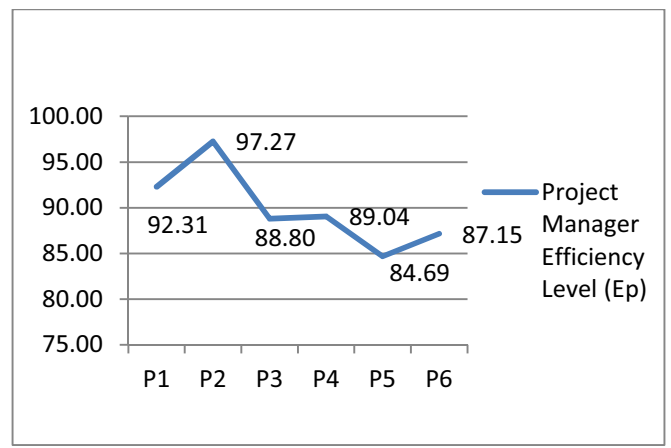

Figure 7. Project Manager Efficiency (Ep) Level for the sampled projects

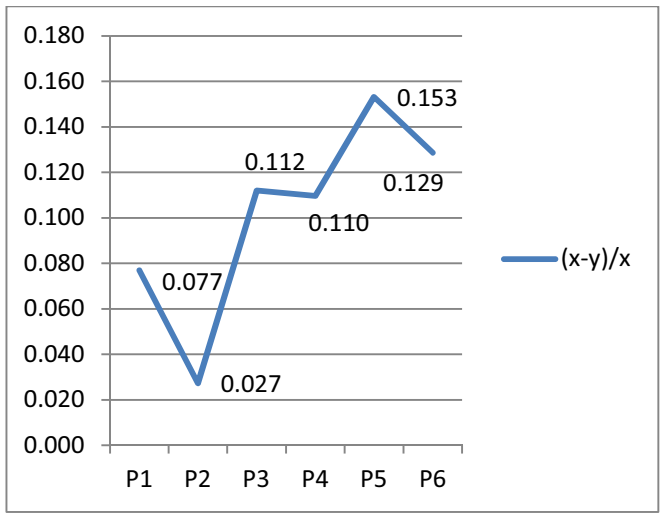

Figure 8. Prediction Efficiency for a Project Manager

This paper creates awareness for the project manager to enhance their efficiency level through the analytical approach. However, this paper does not cover the other coefficient factors as discussed in the above equations.

## VI  CONCLUSION

Effective defect management is one of the influencing factors for success in any software project. Defect estimation and prediction are important areas for effective defect management. The success of the role of a Project Manager depends much on his estimation capability of parameters such as the number of defects and its presence at various phases of software development. Hence, evaluating a project manager analytically becomes very significant in achieving good software quality.

Existing scenario in IT industry indicates that the project manager functions mostly on intuitive analysis. However, since project manager has a vital role in success of the project, it is very important for the project managers to operate using analytical mode of operations.

This paper presents a case study of a product based software industry. The analysis of several projects brings visibility of the role of project managers in achieving effective defect management through computations. The work calls for project managers to enhance their efficiency level through analytical approach in lieu of intuitive strategy of functioning.

## ACKNOWLEDGEMENT

The authors would like to acknowledge the software company involved in this study and the project managers for their invaluable help in providing necessary information for our work under the framework of the Non-Disclosure Agreement.